\documentclass[aps,12pt,nofootinbib,showpacs]{revtex4}
\usepackage{epsf}

\begin{document}
\title{A gluon condensate term in a heavy quark mass}
\author{V.V.Kiselev}
\email{kiselev@th1.ihep.su}
\affiliation{Russian State Research Center "Institute for High
Energy Physics",
Protvino, Moscow Region,  142281,  Russia. \\
 Fax: +7-0967744739}

\begin{abstract}
We investigate a connection between a renormalon ambiguity of
heavy quark mass and the gluon condensate contribution into the
quark dispersion law related with a virtuality defining a
displacement of the heavy quark from the perturbative mass-shell,
which happens inside a hadron.
\end{abstract}


\maketitle
An Operator Product Expansion (OPE) is among the most powerful
tools in the heavy quark physics. In this respect it is usually
applied in the form of series in the inverse heavy quark mass,
determining the characteristic energy scale, say, in sum rules or
for decays etc. \cite{1}. It is well recognized that the Wilson
coefficients standing in front of quark-gluon operators can
contain the uncertainty caused by the factorization of
perturbative contribution and the nonperturbative matrix elements
of composite operators. In this case the restriction on internal
virtualities in Feynman diagrams has to be introduced to control
the dependence on an ``infrared'' energy scale $\lambda$. Usually,
the gluon propagator is modified by replacement: $1/k^2\to
1/(k^2-\lambda_g^2)$ or the cut off the gluon momenta is performed
as $k^2>\lambda^2$ \cite{2}. The calculation results depend on
these parameters. Say, a peculiar behaviour at $\lambda_g^2\to 0$
appears in physical quantities. For example, a perturbative
correlator of two heavy quark currents acquires a power correction
like $\lambda^4/m^4$, where $m$ is the heavy quark mass \cite{3}.
Physically, it means that the OPE can be valid if we sum the
perturbative and nonperturbative parts with the vacuum expectation
of gluon operator which has the same low energy scale dependence:
the gluon condensate $\sim \lambda^4$. Then the
$\lambda$-dependent term can be adopted by an appropriate
definition of OPE with the condensates. Another case takes place
for the uncertainty in the heavy quark mass, where the
perturbative calculation of self-energy with the gluon virtuality
cut off leads to the linear term in $\lambda$. However, there is
no appropriate operator whose vacuum expectation is proportional
to the first power of low energy scale \cite{1}. It was shown that
the mentioned uncertainty proportional to the powers of
factorization scale $\lambda$ can be related with the perturbative
summation of higher order diagrams, which in the limit of
infinitely large number of flavors has the divergency of series in
$\beta_0\alpha_s$, where $\beta_0$ denotes the first coefficient
of Gell-Mann--Low function in QCD. The Borel transform of such
series has some peculiar points, which provide the uncertainty in
the inverse transformation. This uncertainty, related with the
divergency of perturbative series is called the renormalon
\cite{4}, since the physical contents of such fact is clarified by
the representation, where the series are combined in the running
of QCD coupling constant dependent of the gluon virtuality. The
coupling has the singularity, which is the indication of
confinement. In this way, the uncertainty in powers of
$\Lambda_{QCD}$ appears again. Modern studies on the renormalon
applications can be found in \cite{Kata}. These facts imply that
the OPE for fixed values of physical quantities (say, partial
widths or coupling constants in the sum rules) in terms of
perturbative heavy quark mass results in the heavy quark mas,
whose value extracted form the data, strongly depends on the order
of calculation in $\alpha_s$-series \cite{1}: the mass value is
significantly changed from order to order.

Thus, the heavy quark quantities have the renormalon uncertainties connected
to the infrared confinement in QCD. Some of them can be eaten by the
appropriate definition of OPE with condensates. The heavy quark mass is of a
special interest, since its infrared uncertainty cannot be
straightforwardly adopted by the vacuum expectation of an operator with the
dimension 1 in the energy scale.

In present paper we evaluate the gluon condensate contribution to the
dispersion law of heavy quark. We find that the corresponding operator is
divided by the third power of quark virtuality, which results in the
appropriate dimension of term in the heavy quark mass. We discuss how this fact
can be used to cancel the infrared uncertainty of mass.

We perform the calculation of diagram shown in Fig.1 in the technique of
fixed-point gauge \cite{5} with the NRQCD propagators of heavy quarks \cite{6}.

\setlength{\unitlength}{1mm}
\begin{figure}[th]
\begin{center}
\begin{picture}(100,40)
\put(5,0){
\epsfxsize=8cm \epsfbox{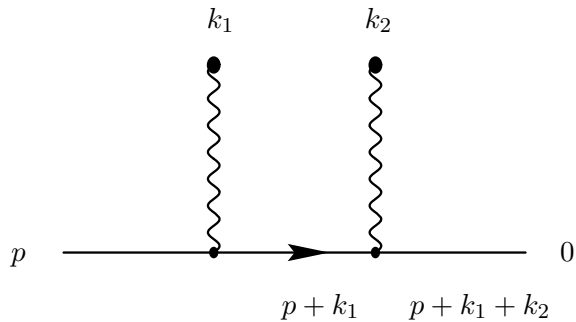}}
\put(7,9){$p$}
\put(80,9){$0$}
\put(43,2){$p+k_1$}
\put(60,2){$p+k_1+k_2$}
\put(33,40){$k_1$}
\put(54,40){$k_2$}
\end{picture}
\end{center}
\caption{The diagram with the gluon condensate contribution to the two-point
effective action of heavy quark.}
\end{figure}

The covariant form of two-point heavy quark effective action $\bar h_v \Gamma
h_v$ can be represented as
\begin{eqnarray}
\Gamma &=& p\cdot v - \frac{(p\cdot v)^2-p^2}{2 m}+\nonumber \\
&&\frac{\pi^2}{24}\left\langle
\frac{\alpha_s}{\pi} G^2_{\mu\nu}\right \rangle \left [ \frac{(p\cdot
v)^2-p^2}{m^2}\frac{1}{\left(p\cdot v - \frac{(p\cdot v)^2-p^2}{2
m}\right)^3}+\frac{1}{m} \frac{1}{\left(p\cdot v - \frac{(p\cdot v)^2-p^2}{2
m}\right)^2}\right],
\label{main}
\end{eqnarray}
where $v$ denotes the four-velocity of hadron containing the heavy quark.
The validity of (\ref{main}) holds under the certain condition on the region of
kinematical variables: the gluon condensate term in the dispersion law of quark
is less than the leading contribution.

In the rest frame of hadron $v=(1,{\bf 0})$ we have
$$
p\cdot v - \frac{(p\cdot v)^2-p^2}{2 m} = p_0 -\frac{{\bf p}^2}{2 m} =
\Delta E,
$$
where $\Delta E$ denotes a heavy quark virtuality inside the hadron. The
perturbative mass-shell is defined by the following expression:
$$
\Delta E = 0.
$$
It is quite clear that the confined quark cannot reach the
mass-shell and there is a minimal displacement from the surface of
free quark motion, which is a nonperturbative quantity. So, we
suppose that
$$
\Delta E \sim \Lambda_{QCD}.
$$
In what follows we apply the model with the quark dispersion law determined by
the form dictated by the account of gluon condensate in (\ref{main}):
\begin{equation}
p_0 = \omega_0+\frac{{\bf p}^2}{2 \tilde m},
\end{equation}
where again $\omega_0\sim \Lambda_{QCD}$ and $\tilde m$ denotes the effective
heavy quark mass, which differs from the perturbative pole mass due to the
contribution of gluon condensate. In the nonrelativistic rest frame we
have\footnote{In NRQCD, where $|{\bf p}|/m < 1$, the gluon condensate
correction to the heavy quark action $\Gamma$ tends to zero at large
virtualities $Q=\Delta E$ as $O(1/Q^2)$ and $O(1/Q^3)$ for the static and
dynamic terms, respectively. However, the correction remains small even at
lower scales.}
\begin{equation}
\Gamma = p_0 - \frac{{\bf p}^2}{2 m}+\frac{\pi^2}{24}\left\langle
\frac{\alpha_s}{\pi} G^2_{\mu\nu}\right \rangle \left [ \frac{{\bf
p}^2}{m^2\Delta E^3}+\frac{1}{m \Delta E^2}\right].
\label{main2}
\end{equation}
Then, we can derive that
\begin{equation}
\tilde m = m + \frac{\pi^2}{12}\left\langle
\frac{\alpha_s}{\pi} G^2_{\mu\nu}\right \rangle \frac{1}{\Delta E^3}.
\label{mm}
\end{equation}
Eq.(\ref{mm}) shows that at $\langle \frac{\alpha_s}{\pi}
G^2_{\mu\nu} \rangle\sim \Lambda_{QCD}^4$ the contribution of
gluon condensate to the heavy quark mass is about $\Lambda_{QCD}$,
i.e. it is linear in the infrared scale of energy, when the
operator determining this term is of the fourth power in the
scale.

Note, that the second term independent of ${\bf p}^2$ in the gluon
condensate contribution shown in (\ref{main2}) results in the
correction to the static energy of heavy quark, so that
\begin{equation}
\delta \omega_0 = \frac{\pi^2}{24}\left\langle
\frac{\alpha_s}{\pi} G^2_{\mu\nu}\right \rangle \frac{1}{m \Delta E^2}.
\label{small}
\end{equation}
Furthermore, the gluon condensate contributes to $\omega_0$ in two
ways: the first one is explicitly given by (\ref{small}), the
second is related with the redefinition of heavy quark mass ($m\to
\tilde m$). Indeed, in this case we have to redefine the ``large''
momentum of heavy quark by the substitution for $m v$ by $\tilde m
v$ and so on, which means that the resulting change of static
energy is given by
$$
\Delta \omega_0 = \tilde m - m +\delta\omega_0 \sim
\Lambda_{QCD}\left(1+\kappa  \frac{\Lambda_{QCD}}{2m}\right),
\;\;\; \kappa \sim 1.
$$
Then, we can see that after the account for the gluon condensate the
displacement of static energy can be basically adopted in the mass $\tilde m$.

Furthermore, we can write down the following relations for the perturbative
dependence of heavy quark quantities on the scale $\lambda$:
\begin{equation}
\frac{d m^{\rm pert}}{d\lambda} = \frac{d \omega_0}{d\lambda} =
\frac{d\Delta E}{d\lambda}, \label{pert}
\end{equation}
where in the second equality we neglect the dynamical term and remain the
static energy. Then the linear dependence on $\lambda$ in $m$ appears in to
ways: the first is the direct calculation of self-energy diagram for the heavy
quark, which results in
$$
\frac{d m^{(1)}}{d\lambda} = C_m \alpha_s(\lambda),
$$
and the second is contributing from the gluon condensate term due to the
$\Delta E$ dependence according to (\ref{mm}) and (\ref{pert}) (the vacuum
condensate of gluon operator has the higher power: $\lambda^4$), so that
$$
\frac{d m^{(2)}}{d\lambda} = - \frac{\pi^2}{4}\left\langle
\frac{\alpha_s}{\pi} G^2_{\mu\nu}\right \rangle \frac{1}{\Delta
E^4}\; C_m \alpha_s(\lambda).
$$
Then, we see that at $\Delta E \approx \omega_0$ the heavy quark mass can be
physically independent on the introduction of factorization scale $\lambda$,
i.e. $\frac{d m}{d\lambda}=\frac{d m^{(1)}}{d\lambda}+\frac{d
m^{(2)}}{d\lambda}=0$, if
$$
\omega_0^4 = \frac{\pi^2}{4}\left\langle
\frac{\alpha_s}{\pi} G^2_{\mu\nu}\right \rangle.
$$
At $\langle \frac{\alpha_s}{\pi} G^2_{\mu\nu} \rangle\approx (0.37\;{\rm
GeV})^4$ \cite{7} the evaluation gives
$$
\omega_0\approx 0.46\;{\rm GeV.}
$$
Neglecting the dynamical term in the heavy quark virtuality we obtain the
following estimate of displacement for the heavy quark mass due to the gluon
condensate:
\begin{equation}
\Delta m \approx \frac{1}{3}\, \omega_0 \approx 0.15\; {\rm GeV,}
\label{strict}
\end{equation}
which can serve as the constrain of maximal value, since we expose
the minimal virtuality.

Thus, the main statement on the nonperturbative displacement of
heavy quark masses remains the following: it is about the
confinement scale. However, we can get some definite estimates for
these values.

The validity of above consideration is determined by the condition
for the formation of hadron containing the heavy quark. Indeed,
the time for the binding of the heavy quark, i.e. for the
formation of its wavefunction, in general depends on the hadron
contents. So, in the heavy-light hadron $H_Q$ with a single heavy
quark, the static energy for light degrees of freedom is of the
order of $\Lambda_{QCD}$, and we get the estimate
$$
\tau[H_Q]\sim \frac{1}{\Lambda_{QCD}},
$$
which is comparable with the characteristic time for the heavy
quark interaction with the gluon condensate. In the doubly heavy
hadron $H_{QQ}$, the formation of wavefunction is determined by
the average size of the doubly heavy system divided by the heavy
quark velocity
$$
\tau[H_{QQ}]\sim \frac{r_{QQ}}{v_Q}\sim\frac{1}{m_Q v_Q^2},
$$
and it depends on the inverse kinetic energy in the doubly heavy
subsystem. So, the calculated contribution by the gluon condensate
in the heavy quark mass would be inapplicable for the quarks
heavier than 20 GeV. However, in the systems composed by charmed
and beauty quarks the kinetic energy is about $\Lambda_{QCD}$, and
in the reality we deal with the situation, when the effects
connected with the formation of wavefunction for the hadron
containing the heavy quark and the gluon condensate term are
competitive. So, for instance, the energy shift due to the
interaction of coulomb doubly heavy system with the gluon
condensate \cite{MV} is determined by the following expression:
\begin{equation}
\Delta E_{Q\bar Q} = \frac{\pi^2}{18}\,\left\langle
\frac{\alpha_s}{\pi} G^2_{\mu\nu}\right \rangle\,
\frac{n^2m}{(mE_n)^2}\, \epsilon_{nl},\qquad E_n =
-\frac{1}{4n^2}\,\left(\frac{2}{3}\,\alpha_s\right)^2\, m,
\label{MV}
\end{equation}
where $\epsilon_{nl}$ is a rational factor depending on the
principal and radial quantum numbers $n$ and $l$. Comparing
(\ref{MV}) with (\ref{mm}), we see that despite different
approaches these two equations can be in agreement with each
other, if we substitute of $\Delta E \sim E_n$, i.e. if the
virtuality is determined by the bound energy of the heavy quark in
the heavy quarkonium system. This fact implies that the heavy
quarkonium represents a specific case of the general consideration
applied to the coulomb system, wherein the virtuality is
prescribed to a concrete value. As for the numerical estimates, we
have to take into account, that one should use (\ref{mm}) instead
of (\ref{MV}), if the virtuality of the heavy quark in the
quarkonium is less that the value following from the general form
of dispersion law for the quark, i.e. it is less than $0.46$ GeV.
Otherwise, the quark is heavy enough in order to use the coulomb
approximation of (\ref{MV}).

To conclude, we have shown that the Operator Product Expansion including the
gluon condensate results in the following dispersion law for the heavy quark:
$$
p_0({\bf p}) = \omega_0 + \frac{{\bf p}^2}{2 m},
$$
where the correction to the heavy quark mass is given by
$$
\Delta m = \frac{\pi^2}{12}\left\langle
\frac{\alpha_s}{\pi} G^2_{\mu\nu}\right \rangle \frac{1}{\omega_0^3},
$$
and the infrared ambiguity in the mass caused by the corresponding renormalon,
can be cancelled at
$$
\omega_0^4 = \frac{\pi^2}{4}\left\langle
\frac{\alpha_s}{\pi} G^2_{\mu\nu}\right \rangle.
$$
Of course, the conclusion is drawn to the given, linear order in $\alpha_s$,
and the well known divergency of heavy quark pole mass with the increase of
$\alpha_s$-order probably can be removed, if the higher order corrections to
the Wilson coefficient of gluon condensate as well as the higher condensates
will be included into the consideration in the same manner.

The author expresses the gratitude to A.L.Kataev for fruitful
discussions and valuable remarks.

This work is in part supported by the Russian Foundation for Basic Research,
grants 99-02-16558 and 96-15-96575.

\end{document}